\documentclass[conference]{IEEEtran}
\IEEEoverridecommandlockouts

\usepackage{graphicx}
\usepackage{textcomp}
\usepackage{xcolor}

\usepackage[letterpaper, total={180mm,229mm}]{geometry}
\def\BibTeX{{\rm B\kern-.05em{\sc i\kern-.025em b}\kern-.08em
    T\kern-.1667em\lower.7ex\hbox{E}\kern-.125emX}}

\usepackage{graphicx}
\usepackage{amsmath,amssymb,amsfonts}
\usepackage[noend,linesnumbered,ruled,vlined]{algorithm2e}
\usepackage{booktabs}
\usepackage{listings}
\usepackage{microtype}
\usepackage{subcaption}
\usepackage{multirow}
\usepackage{amsmath}
\usepackage{amsthm}
\usepackage{hyperref}
\usepackage{tabularx}
\usepackage{pgfplots}

\hypersetup{hidelinks}

\usepackage[edges]{forest}
\forestset{
  dir tree/.style={
    for tree={
      parent anchor=south west,
      child anchor=west,
      anchor=mid west,
      inner ysep=-1pt,
      grow'=0,
      align=left,
      edge path={
        \noexpand\path [draw, \forestoption{edge}] (!u.parent anchor) ++(0.5em,-0.5em) |- (.child anchor)\forestoption{edge label};
      },
      font=\footnotesize,
      if n children=0{}{
        delay={
          prepend={[,phantom, calign with current]}
        }
      },
      fit=band,
      before computing xy={
        l=2em
      }
    },
  }
}

\renewcommand{\footnoterule}{%
	\kern -3pt
	\hrule
	\kern 2pt
}

\usepackage{microtype}
\usepackage{rotating}
\usepackage{adjustbox}
\SetCommentSty{mycommfont}
\usepackage[style=ieee, sorting=none, mincitenames=1,backend=bibtex,doi=false,isbn=false,url=false]{biblatex} %
\usepackage{scalerel}
\usepackage{tikz} 

\usetikzlibrary{svg.path}

\definecolor{orcidlogocol}{HTML}{A6CE39}
\tikzset{
  orcidlogo/.pic={
    \fill[orcidlogocol] svg{M256,128c0,70.7-57.3,128-128,128C57.3,256,0,198.7,0,128C0,57.3,57.3,0,128,0C198.7,0,256,57.3,256,128z};
    \fill[white] svg{M86.3,186.2H70.9V79.1h15.4v48.4V186.2z}
                 svg{M108.9,79.1h41.6c39.6,0,57,28.3,57,53.6c0,27.5-21.5,53.6-56.8,53.6h-41.8V79.1z M124.3,172.4h24.5c34.9,0,42.9-26.5,42.9-39.7c0-21.5-13.7-39.7-43.7-39.7h-23.7V172.4z}
                 svg{M88.7,56.8c0,5.5-4.5,10.1-10.1,10.1c-5.6,0-10.1-4.6-10.1-10.1c0-5.6,4.5-10.1,10.1-10.1C84.2,46.7,88.7,51.3,88.7,56.8z};
  }
}

\newcommand\orcidicon[1]{\href{https://orcid.org/#1}{\mbox{\scalerel*{
\begin{tikzpicture}[yscale=-1,transform shape]
\pic{orcidlogo};
\end{tikzpicture}
}{|}}}}

\begin{document}

\onecolumn
\thispagestyle{empty}
\twocolumn
\setcounter{page}{1}
\setcounter{figure}{0}

\title{\textit{GRAU}: A Generic Runtime-Reconfigurable Activation–Requantization Unit for Quantized Neural Network Accelerators
}

\author{
\IEEEauthorblockN{Yuhao Liu$^{1,2,3}$ \orcidicon{0000-0002-7281-2126}, \textit{Member, IEEE}, Salim Ullah$^{1}$ \orcidicon{0000-0002-9774-9522}, Akash Kumar$^{1}$ \orcidicon{0000-0001-7125-1737}, \textit{Senior Member, IEEE}}
\IEEEauthorblockA{$^{1}$Ruhr University Bochum, Germany $^{2}$Dresden University of Technology, Germany\\
$^{3}$Center for Scalable Data Analytics and Artificial Intelligence (ScaDS.AI Dresden/Leipzig), Germany\\
Email: \{yuhao.liu, salim.ullah, akash.kumar\}@rub.de}
}


\maketitle

\begin{abstract}
		 As neural networks continue to scale, low-precision quantization is widely adopted in edge accelerators, but conventional multi-threshold activation hardware requires $2^{n}-1$ thresholds for $n$-bit integer outputs, leading to rapidly increasing hardware costs. We propose GRAU, a \emph{Generic Runtime-Reconfigurable Activation-Requantization Unit} that fuses batch normalization, nonlinear activation, and requantization through piecewise-linear fitting with \emph{Power-of-Two} slopes. GRAU requires only basic comparators and 1-bit right shifters, while supporting mixed-precision quantization and runtime switching among post-accumulation functions, including modern nonlinear activations such as \emph{SiLU} and \emph{GELU}. Compared with multi-threshold activation-requantization units, GRAU reduces LUT consumption by approximately $90\%$ or more while improving hardware efficiency, flexibility, and scalability. We further explore the trade-off between accuracy and hardware cost across different approximation settings and provide an open-source toolchain for GRAU-aware network training.

\end{abstract}%

\thispagestyle{empty}


\section{Introduction}
\label{introduction}
\begin{figure}[t]
    \centering
    \centerline{\includegraphics[width=1\columnwidth]{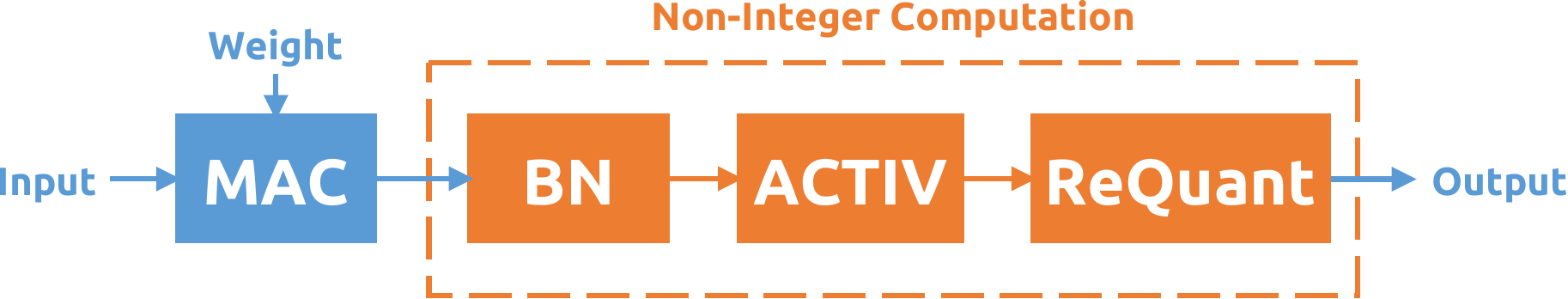}}
    \caption{Non-Integer Computation in Quantized Neural Network}
    \label{non_int}
\end{figure}

Continuously growing sizes of state-of-the-art neural network models encourage researchers to explore different schemes to accelerate network inference and improve power efficiency by reducing memory consumption and computing power requirements. Therefore, quantization becomes one of the most widely applied methods in related works, especially by training the weights and activations in the network model as low-precision integers. Considering that the outputs of each neuron or kernel in~\emph{Quantized Neural Networks} (QNNs) should be quantized integers following the selected precision, one re-quantization unit should be implemented after the activation unit to convert the outputs of the activation function to integers in the design of a QNN hardware accelerator. Therefore, as shown in~\autoref{non_int}, \emph{Batch Normalization}, nonlinear activation, and re-quantization are the non-integer computation parts in the QNN inference. 

\begingroup
    \renewcommand{\thefootnote}{}
    \footnotetext{Accepted to special session track}
\endgroup

\subsection{Motivation}
Because the large multiplier and nonlinear computation in the orange block shown in~\autoref{non_int} are expensive on hardware implementation, previous works have extensively explored the designs of related quantized units in QNN accelerators. One of the widely used design paradigms is the \emph{Multi-Threshold} (MT) activation unit, adopted in well-known designs such as \emph{FINN}~\cite{Umuroglu2017} and \emph{FINN-R}~\cite{Blott2018}. By folding \emph{Batch Normalization}, nonlinear activation, and re-quantization into a single comparator-based block, MT units replace expensive arithmetic operations with a set of fixed thresholds. However, this design paradigm faces three major limitations as QNNs evolve:

\begin{figure*}[t]
    \centering
    \begin{minipage}{0.94\columnwidth}
        \centerline
        {
            \resizebox{\columnwidth}{!}
            {
                \begin{tikzpicture}[x=1mm,y=1mm]
                    \begin{axis}[
                        width=10cm,
                        height=5cm,
                        scale only axis,
                        axis lines=middle,
                        xlabel=$x$,
                        ylabel=$\sigma(x)$,
                        samples=100,
                        domain=-6:6,
                        ymin=-0, ymax=1.0,
                        xmin=-6, xmax=6,
                        grid=major,
                        enlargelimits=true,
                        axis on top,
                    ]




                        
                        \addplot[draw={rgb,255:red,192; green,0; blue,0}, line width=1.5pt] {1 / (1 + exp(-x))};

                        \foreach \x in {-2,1,4} 
                        {

                            \addplot[draw={rgb,255:red,31; green,78; blue,121}, dashed, line width=1.5pt] coordinates {(\x,0) (\x,{1 / (1 + exp(-\x))})};

                            \addplot[draw={rgb,255:red,31; green,78; blue,121}, dashed, line width=1.5pt] coordinates {(\x,{1 / (1 + exp(-\x))}) (0,{1 / (1 + exp(-\x))})};
                        }

                        \node at (axis cs:0,{1 / (1 + exp(2))}) [anchor=west, text={rgb,255:red,31; green,78; blue,121}, font=\large\bfseries] {1};
                        \node at (axis cs:0,{1 / (1 + exp(-1)) + 0.05}) [anchor=west, text={rgb,255:red,31; green,78; blue,121}, font=\large\bfseries] {2};
                        \node at (axis cs:0,{1 / (1 + exp(-4)) - 0.05}) [anchor=west, text={rgb,255:red,31; green,78; blue,121}, font=\large\bfseries] {3}; 

                    \end{axis}
                \end{tikzpicture}
            }
        }
    \end{minipage}
    \begin{minipage}{\columnwidth}
        \centerline
        {
            \resizebox{\columnwidth}{!}
            {
                \resizebox{\columnwidth}{!}
                {
                    \begin{tikzpicture} [x=1mm,y=1mm]
                        \begin{axis}[
                            width=10cm,
                            height=5cm,
                            scale only axis,
                            axis lines=middle,
                            xlabel=$x$,
                            ylabel=$G(0.25x-0.5)$,
                            samples=100,
                            domain=-8:13,
                            ymin=-0.0, ymax=1.0,
                            xmin=-8, xmax=13,
                            grid=major,
                            enlargelimits=true,
                            axis on top,
                        ]

    


                            \addplot[draw={rgb,255:red,192; green,0; blue,0}, line width=1.5pt] {exp(-1 * (0.25 * (x-2))^2)};

                            \foreach \x in {-4,1,6} 
                            {
                                \addplot[draw={rgb,255:red,31; green,78; blue,121}, dashed, line width=1.5pt] coordinates {(\x,0) (\x,{exp(-1 * (0.25 * (\x-2))^2)})};
    
                                \addplot[draw={rgb,255:red,31; green,78; blue,121}, dashed, line width=1.5pt] coordinates {(\x,{exp(-1 * (0.25 * (\x-2))^2)}) (0,{exp(-1 * (0.25 * (\x-2))^2)})};
                            }

                            \node at (axis cs:0,{exp(-1 * (0.25 * (-4-2))^2)-0.05}) [anchor=east, text={rgb,255:red,31; green,78; blue,121}, font=\large\bfseries] {1};
                            \node at (axis cs:0,{exp(-1 * (0.25 * (1-2))^2)-0.01}) [anchor=east, text={rgb,255:red,31; green,78; blue,121}, font=\large\bfseries] {3?};
                            \node at (axis cs:0,{exp(-1 * (0.25 * (6-2))^2)-0.05}) [anchor=east, text={rgb,255:red,31; green,78; blue,121}, font=\large\bfseries] {2?}; 
                        \end{axis}
                    \end{tikzpicture}
                }
            }
        }
    \end{minipage}
    \caption{Correct 2-bit quantization of \emph{Multi-Threshold} unit (left) in \emph{Sigmoid} and the mistake of \emph{Multi-Threshold} unit in non-monotonically increasing function (right)}
    \label{multi-threshold-plot}
\end{figure*}
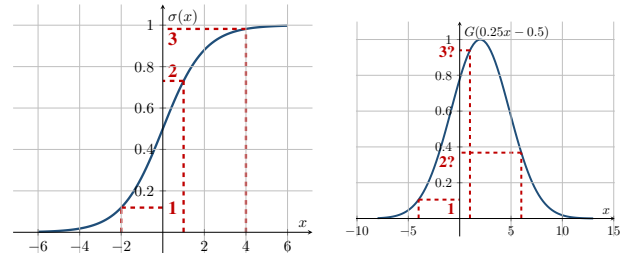

\subsubsection{Exponential hardware scaling with precision}

MT unit takes the integer outputs from \emph{Multiply-Accumulators} (MAC) and compares them with $2^n-1$ thresholds to produce $n$-bit outputs, such as 15 thresholds for 4-bit and 255 for 8-bit. If the MAC result exceeds $m$ thresholds, the quantized activation output is an integer, $m$. This design paradigm suggests exponentially increasing hardware resource consumption, since following an increase in output precision, the number of thresholds grows exponentially. 

\begin{table}[t]
    \centering
    \caption{Comparison between Unified-Precision and Mixed-Precision QNN on MNIST~\cite{bitsys2}}
    \vspace{-5pt}
    \resizebox{1\columnwidth}{!}
    {
        \begin{tabular}{cccccc}
            \toprule
            \multirow{2}{*}{Model} & \multirow{2}{*}{Precision} & Accu. & $\Delta$Accu. & Memory  & Baseline \\
                                   &                            & (\%)  & (pp)          & (Bytes) & Ratio    \\ \midrule
            \multirow{3}{*}{MLP}   & Full 1-bit                 & 92.29 & -3.62         & 7376    & 0.74     \\
                                   & Mixed (baseline)           & 95.91 & 0.00          & 9984    & 1.00     \\
                                   & Full 8-bit                 & 97.36 & 1.45          & 59008   & 5.91     \\ \midrule
            \multirow{3}{*}{CNN}   & Full 1-bit                 & 96.26 & -2.53         & 29848   & 0.54     \\
                                   & Mixed (baseline)           & 98.79 & 0.00          & 55712   & 1.00     \\
                                   & Full 8-bit                 & 99.14 & 0.35          & 238784  & 4.29     \\ \bottomrule
        \end{tabular}
    }
    \label{mixed_precision_tab}
\end{table}

\subsubsection{Inefficiency in mixed-precision quantization}

Mixed-precision quantization has recently emerged as a key trend in the design of lightweight AI on the edge. For instance, Liu et al.~\cite{bitsys2} compared unified-precision and mixed-precision QNNs using a small 4-layer MLP and CNN on MNIST~\cite{deng2012mnist}. Their results in~\autoref{mixed_precision_tab} indicate that 1/2/4/8-bit mixed precision offers a trade-off between accuracy and memory usage compared to BNN and QNN, tolerating a slight accuracy loss to significantly reduce memory cost compared to QNN. However, the MT unit must implement the maximum number of thresholds required by the highest precision. For instance, in 1/2/4/8-bit mixed-precision quantization, the MT unit must implement 255 thresholds for the 8-bit precision, while only a small subset is used for lower precisions (for example, only one threshold is used for 1-bit). Serial reuse of a comparator can reduce hardware costs but significantly increases latency at higher precisions. Therefore, implementing how many MT units on a QNN accelerator, especially for the edge device, becomes a critical design trade-off between resource consumption and inference latency.

\subsubsection{Inability to represent non-monotonic activations}
Since MT outputs always increase as more thresholds are exceeded by inputs, the design inherently supports only monotonically increasing functions. Emerging nonlinear activations, such as \emph{SiLU}~\cite{SiLU}, violate this constraint, making the MT unit incompatible with many practical QNN settings. The left plot in~\autoref{multi-threshold-plot} instantiates a correct quantization processing of a \emph{Sigmoid} function with three thresholds for 2-bit output. However, the right plot shows an incompatible case of MT units where the expected output, 3, at the threshold of 0.9, is larger than the expected output, 2, at 6.0, violating the monotonically increasing condition. The quantized output at 6.0 should be 3 since it exceeds three thresholds, and the output at 0.9 should be 2 instead.

\subsection{Related Works}

Although the MT unit has been successfully adopted in various well-known works, considering the above-discussed shortcomings, we further surveyed other potential solutions introduced in prior works, such as \emph{Piecewise Linearized Fitting} (PWLF), \emph{Second-Order Polynomial Fitting} (SOPF), and \emph{Look-up Table} (LUT):

\begin{itemize}

    \item PWLF and SOPF are more flexible than MT units to fit various nonlinear functions, such as: i) ML-PLAC~\cite{ml-plac} explored the Power-of-Two-based PWLF approximation for nonlinear functions, like \emph{Tanh}, \emph{Softsign}, \emph{$2^x$}, \emph{$log_2(1+x)$}, etc. ii) Li et al.~\cite{Li2022}, Tsmots et al.~\cite{Tsmots2019}, and Nguyen et al.~\cite{Nguyen2020} explored the PWLF-based hardware design of~\emph{Sigmoid} activation functions. iii) Liu et al.\cite{Liu2023_2} and Bouguezzi et al.~\cite{Bouguezzi} presented the PWLF-based implementation for \emph{Tanh} and \emph{TanhExp}. iv) Tsmots et al. and Bouguezzi et al. applied the SOPF scheme for \emph{Sigmoid} and \emph{Tanh} approximation on hardware. However, these schemes are primarily focused on recovering the accuracy of the nonlinear function. For example, ML-PLAC~\cite{ml-plac} implements 24 and 42 segments for \emph{Tanh(x)} and \emph{Softsign(x)}, respectively, to recover the accuracy of these nonlinear functions. The \emph{Batch Normalization} and \emph{Output Re-quantization} before and after the nonlinear functions in the QNN accelerator design have not been folded with them to reduce resource consumption. However, according to the theory of approximation computing, neural networks can tolerate computational errors, such as approximations in multiplication or in activation functions. Therefore, allocating more hardware resources to recover higher accuracy of nonlinear activations may not be the most hardware-efficient choice for resource-constrained QNN accelerators.  
    
    \item Compared to the \emph{Multi-Threshold} activation method, the Lookup-Table-based quantized activation unit is also hardware-friendly, as shown in the works of Piazza et al.~\cite{Piazza1993}, Pogiri et al.~\cite{Pogiri2022}, and Kumar Meher et al.~\cite{Kumar2010}. In principle, by rewriting the lookup contents stored in flip-flops or BRAM, the activation function can be reconfigured at runtime. However, such schemes are less suitable for a generic runtime-reconfigurable activation unit like the MT activation unit. For a direct lookup implementation, the storage cost grows exponentially with the input address width and linearly with the output bit-width. In our experiments, the integer MAC outputs of 8-bit quantized ResNet-18~\cite{resnet} on ImageNet~\cite{imagenet-dataset} can reach approximately $[-10^5,10^5]$, which already corresponds to an address space close to 18 bits. Mapping such inputs directly to 8-bit outputs (e.g., $[-128,127]$) would require a prohibitively large lookup table. Furthermore, in mixed-precision designs, different precision modes and activation functions require different lookup contents, implying either multiple tables or runtime table rewriting, which further increases memory overhead and control complexity.
    
\end{itemize}

\subsection{Contribution}

\begin{table}[t]
    \centering
    \caption{Comparison between SOPF, PWLF, MT, LUT, and GRAU for Quantized Activation Unit}
    \vspace{-5pt}
    \resizebox{\columnwidth}{!}
    {
        \begin{tabular}{cccccc}
            \toprule
            Method              & SOPF                 & PWLF                 & MT                    & LUT                      & \textbf{GRAU}                  \\ \midrule
            Hardware            & \multirow{2}{*}{Low} & \multirow{2}{*}{Low} & \multirow{2}{*}{High} & \multirow{2}{*}{High}    & \multirow{2}{*}{\textbf{High}} \\
            Friendly            &                      &                      &                       &                          &                       \\ \midrule
            Runtime             & \multirow{2}{*}{No}  & \multirow{2}{*}{No}  & \multirow{2}{*}{Yes}  & \multirow{2}{*}{Limited} & \multirow{2}{*}{\textbf{Yes}}  \\
            Reconfigurable      &                      &                      &                       &                          &                       \\ \midrule
            Non-monotonically   & \multirow{2}{*}{Yes} & \multirow{2}{*}{Yes} & \multirow{2}{*}{No}   & \multirow{2}{*}{Yes}     & \multirow{2}{*}{\textbf{Yes}}  \\
            Increasing Function &                      &                      &                       &                          &                       \\ \midrule
            Adaptive for        & \multirow{2}{*}{Yes} & \multirow{2}{*}{Yes} & \multirow{2}{*}{No}   & \multirow{2}{*}{No}      & \multirow{2}{*}{\textbf{Yes}}  \\
            Mixed-Precision     &                      &                      &                       &                          &                       \\ \midrule
            End-to-End          & \multirow{2}{*}{No}  & \multirow{2}{*}{No}  & \multirow{2}{*}{Yes}  & \multirow{2}{*}{Yes}     & \multirow{2}{*}{\textbf{Yes}}  \\
            MAC to Quant        &                      &                      &                       &                          &                       \\ \bottomrule
        \end{tabular}
    }
    \label{activation_unit_compare}
\end{table}

As summarized in~\autoref{activation_unit_compare}, comparing the four activation hardware design paradigms discussed above, we noticed that the unified, hardware-efficient, and reconfigurable design, which can simultaneously support multi-function, mixed-precision, and non-monotonic activations, for integer-aware QNN accelerators, is still not well supported by existing works. Therefore, we propose the \emph{\textbf{G}eneric \textbf{R}untime-Reconfigurable \textbf{A}ctivation-Requantization \textbf{U}nit} (GRAU) for low-precision quantized, integer-based QNN hardware accelerators with flexible support of multi-activation functions and mixed-precision quantization. Our work does not attempt to implement an approximate hardware generation framework for specific nonlinear functions, like previous works such as ML-PLAC~\cite{ml-plac}. Our design proposes one reusable end-to-end activation processing unit (from MAC output to quantized activated output). The main contributions of this work are as follows:

\begin{itemize}

    \item \textit{Hardware Friendly and Runtime Reconfiguration}: We propose a reusable generic activation unit based on PWLF with \emph{Power-of-Two} (PoT) and \emph{Additive Power-of-Two} (APoT) slope approximation. The design uses only comparators and a 1-bit shifter pipeline and can be reconfigured at runtime by updating a small set of breakpoint and shift-encoding registers. While PWLF, PoT, and APoT techniques have been individually explored in prior work, such as~\cite{ml-plac}, no existing design integrates these concepts into an approximate, reusable, unified activation-requantization processing unit that folds \emph{Batch Normalization} and \emph{Output ReQuantization}, with runtime configurability and mixed-precision integer support.
    
    \item \textit{Flexible Function Support and Adaptation for Mixed-Precision Quantization}: 
    GRAU supports multiple nonlinear and non-monotonic activations (e.g., \emph{ReLU}, \emph{Sigmoid}, \emph{SiLU}~\cite{SiLU}, \emph{GELU}~\cite{hendrycks2016gelu}) and easily adapts to different quantization precisions through lightweight reconfiguration of breakpoints and shifts.
    
    \item \emph{Approximation Experiments}: We developed a fast greedy piecewise linear fitting algorithm to replace the nonlinear activation folded with batch normalization and output re-quantization in Quantized Neural Networks (QNN). We evaluate PoT/APoT approximations on CIFAR-10 and ImageNet with VGG16 and ResNet-18 models under 4/8-bit unified-precision and 2/4/8-bit mixed-precision settings with various activation functions like ReLU, Sigmoid, SiLU,  and GELU. The results suggest the feasibility of our approach for hardware. The greedy piecewise linear fitting algorithm, integrated with PyTorch, has been open-sourced to facilitate easier reproduction of our work.\footnote{https://github.com/liuyh-Horizon/GRAU-Approx-Toolkit}
    
    \item \emph{Hardware Implementation}: We implement both pipelined and serialized GRAU variants. The pipelined design achieves higher throughput, while the serialized version provides lower cost and greater configurability.
    
    \item \emph{Resource Report}: Based on the synthesis and implementation in Vivado, the results show that our GRAU hardware reduces LUT usage by approximately $90\%$ or more compared with \emph{Multi-Threshold} (MT) units, achieving higher frequency, lower \emph{Area-Delay-Product} (ADP), and lower \emph{Power-Delay-Product} (PDP), which demonstrates promising hardware and power efficiency. Our implemented GRAU variants achieve 250 MHz in our Vivado post-implementation results, compared with 200 MHz for the pipelined MT baseline.
    
\end{itemize}

Although GRAU can be integrated into any QNN accelerator on FPGA/ASIC, the goal of this work is not to design a full accelerator architecture. Instead, we aim to establish a unified activation hardware design that enables runtime reconfiguration, multi-function support, and mixed-precision activation with significantly reduced overhead. Since quantized activation functions are present in every layer of modern QNNs, improving each activation unit directly scales across the entire accelerator.

\begin{figure*}[t]
    \centering
    \begin{minipage}{1\columnwidth}
        \centerline{\includegraphics[width=2\columnwidth]{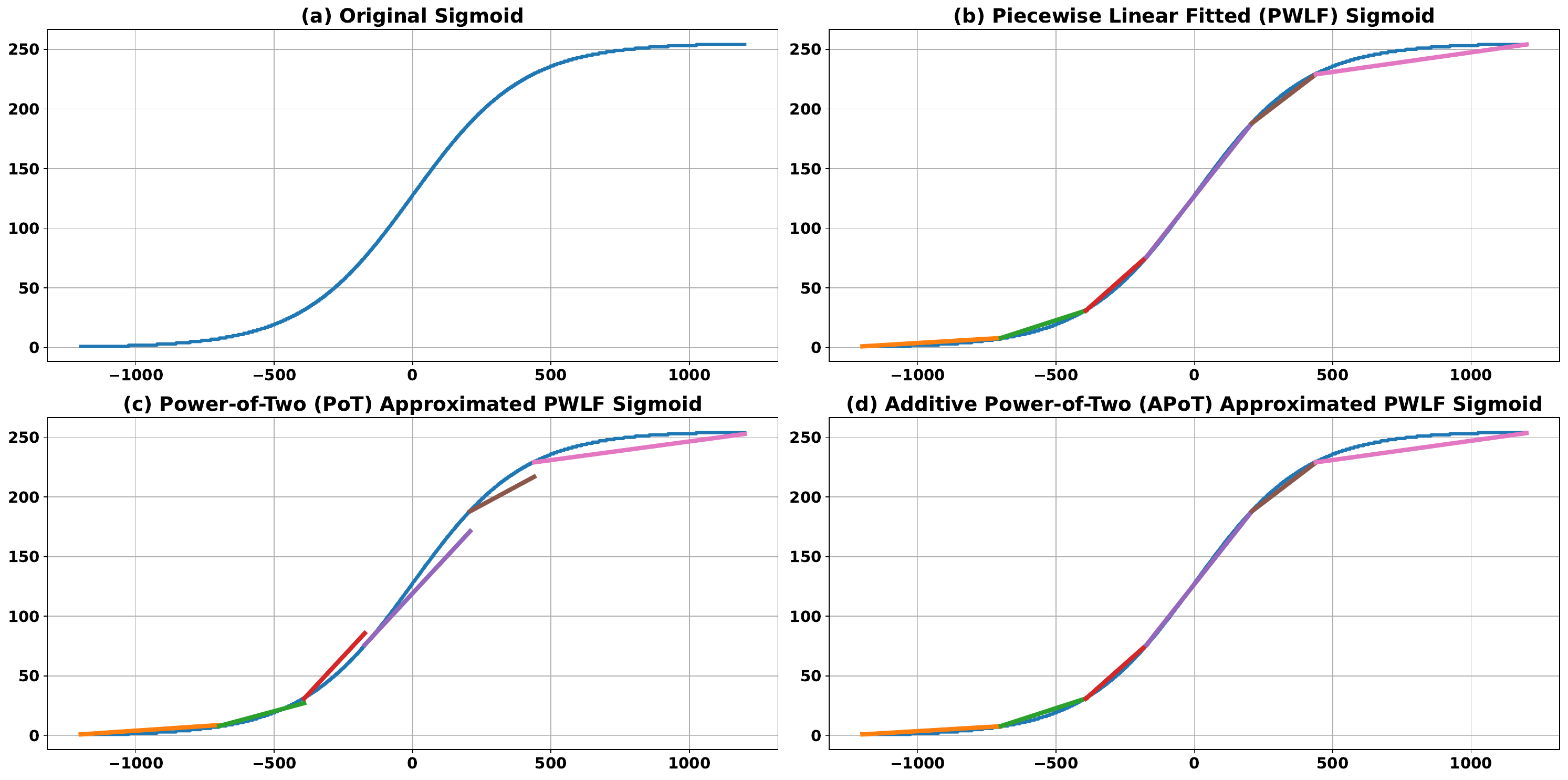}}
    \end{minipage}
    \caption{Comparing the original nonlinear function, PWLF approximated function, PoT approximated PWLF function, and APoT approximated PWLF function}
    \label{pwlf_example}
\end{figure*}

\subsection{Organization}

This manuscript is structured in the following way: \textbf{Section II} discusses how to convert the original QNN models to the PoT and APoT approximated models for GRAU hardware and the hardware designs of GRAU activation units in this work.~\textbf{Section III} shows the hardware evaluation results of the above-mentioned GRAU designs.~\textbf{Section IV} discusses the further potential improvement and optimization of~\emph{GRAU} and concludes the contents of this paper.

\section{Implementation}
\label{imp}
\subsection{PoT and APoT Approximated Activation Functions for GRAU Hardware}


Following the objectives outlined above, we transform the original nonlinear activation functions folded with BN and output re-quantization in QNNs into PWLF, PoT-PWLF, and APoT-PWLF representations compatible with the proposed GRAU architecture. To ensure efficient hardware realization, we additionally constrain the number of segments and limit the allowable range of power-of-two slopes when generating the PoT- and APoT-based approximations.

In~\autoref{pwlf_example}, we introduce an instance of \emph{Sigmoid} functions and their PoT and APoT variants with six segments for 8-bit quantization.~\autoref{pwlf_example}a plots the original \emph{Sigmoid} folded with BN and output re-quantization.~\autoref{pwlf_example}b is its piecewise-linear fitting approximation with 6 segments.~\autoref{pwlf_example}c and~\autoref{pwlf_example}d are its PoT-PWLF and APoT-PWLF approximated functions. 

From PWLF to PoT- and APoT-PWLF, we adopt a three-step approximation: 

\begin{itemize}
    \item Considering that the inputs to our quantized activation unit in the QNN accelerator are the integer outputs from MACs, in the PoT-PWLF and APoT-PWLF approximations, we adjust the breakpoints of segments to their nearest integers. 
    
    \item We approximate the slope of each segment in PWLF functions to the nearest PoT and APoT value. For instance, if we define the allowable power range for PoT and APoT approximation as $[-10, -6]$, it means PoT approximated slopes can be $2^{-10}$, $2^{-9}$, $\dots$, $2^{-7}$, $2^{-6}$, and APoT slopes can be the sum of one combination with any of these allowable PoT values, where one PoT value can only be used once in combination.

    \item We chose the left rounded breaking point of each segment to create a new linear function with approximated PoT and APoT slopes. Therefore, as shown in~\autoref {pwlf_example}c, the PoT approximation has a small gap at the right end of each segment, since the approximated breakpoints and slopes have a small bias for each segment compared with the original PWLF functions. APoT approximation also has this gap. However, it's more accurate than PoT. Therefore, the fourth column cannot clearly show the tiny gap in APoT.
\end{itemize}

Therefore, based on \emph{Brevitas}~\cite{brevitas} and the piecewise linear fitting algorithm, we created the PWLF, PoT-PWLF, and APoT-PWLF approximated QNN models on CIFAR-10 and ImageNet after the \emph{Quantization-Aware Training} (QAT) as the following steps:

\begin{itemize}
    \item Training the QNN models and recording the output ranges of each quantized \emph{Fully Connected} (FC) layer, \emph{QuantLinear}, and quantized \emph{Convolution} (CONV) layers, \emph{QuantConv2d} while training, which are the MAC outputs on hardware. 
    \item For each layer, doubling the recorded MAC output range and averagely generating 1000 samples from the extended range as dummy input. Then, extracting the corresponding BN and quantized activation layers, such as \emph{QuantReLU} and \emph{QuantSigmoid} in \emph{Brevitas}, from the trained QNN model and packaging them as black boxes. Compute the output of these black boxes with the dummy input. Then, adopt these outputs to fit a piecewise linear function based on the piecewise linear fitting algorithm, which will be in the same form as the~\autoref{pwlf_example}b.
    \item Extracting the breakpoints and slopes from the fitted piecewise linear function. Rounding the breakpoints and approximating the slopes in PoT and APoT form with selected power ranges. Then, we created the PoT and APoT PWLF functions and replaced the BN and quantized activation layers in the original model to generate the PWLF, PoT-PWLF, and APoT-PWLF approximated models. 
    \item Evaluating the accuracy of the original accurate model and PWLF, PoT-PWLF, and APoT-PWLF approximated models
\end{itemize}

\begin{table*}[t]
    \centering
    \caption{Design Space Exploration for PoT-PWLF/APoT-PWLF Approximation based on Original QNN and PWLF Models with CIFAR-10~\cite{Krizhevsky2009} and VGG16~\cite{Simonyan2014} Architecture}
    \resizebox{\linewidth}{!}
    {
        \begin{tabular}{cccccccccc}
            \toprule
            \multicolumn{1}{l}{} & \multicolumn{9}{c}{Experiment Result}                                                                                                                                                          \\ \midrule
            Precision            & \multicolumn{3}{c}{4-bit}                                    & \multicolumn{3}{c}{8-bit}                                       & \multicolumn{3}{c}{Mixed-Precision}                            \\
            Activation           & ReLU               & Sigmoid            & SiLU               & ReLU               & Sigmoid             & SiLU                & ReLU               & Sigmoid             & SiLU                \\
            Original             & 91.18\%            & 87.81\%            & 89.55\%            & 92.10\%            & 88.71\%             & 92.35\%             & 92.11\%            & 88.39\%             & 91.10\%             \\ \midrule
            Max. Segment         & \multicolumn{9}{c}{PWLF}                                                                                                                                                                       \\ \midrule
            6                    & 91.00\%            & 86.72\%            & 84.88\%            & 92.08\%            & 76.79\%             & 91.92\%             & 91.78\%            & 80.75\%             & 88.20\%             \\
            8                    & 91.08\%            & 85.99\%            & 85.58\%            & 92.10\%            & 85.87\%             & 92.32\%             & 91.86\%            & 74.87\%             & 88.89\%             \\ \midrule
            Max. Segment         & \multicolumn{9}{c}{PoT-PWLF (Accuracy/Exponent Range)}                                                                                                                                         \\ \midrule
            \multirow{4}{*}{6}   & 90.24\%            & 67.89\%            & 80.29\%            & 91.28\%            & 30.98\%             & 90.81\%             & 91.15\%            & 51.72\%             & 83.12\%             \\
                                 & ($2^{-18}\sim2^{-3}$) & ($2^{-16}\sim2^{-1}$) & ($2^{-16}\sim2^{-1}$) & ($2^{-22}\sim2^{-7}$) & ($2^{-22}\sim2^{-7}$)  & ($2^{-18}\sim2^{-3}$)  & ($2^{-20}\sim2^{-5}$) & ($2^{-16}\sim2^{-1}$)  & ($2^{-20}\sim2^{-5}$)  \\
                                 & 90.24\%            & 64.18\%            & 82.50\%            & 91.28\%            & 27.76\%             & 90.40\%             & 89.89\%            & 17.37\%             & 82.04\%             \\
                                 & ($2^{-10}\sim2^{-3}$) & ($2^{-10}\sim2^{-3}$) & ($2^{-12}\sim2^{-5}$) & ($2^{-14}\sim2^{-7}$) & ($2^{-14}\sim2^{-7}$)  & ($2^{-16}\sim2^{-9}$)  & ($2^{-14}\sim2^{-7}$) & ($2^{-16}\sim2^{-9}$)  & ($2^{-14}\sim2^{-7}$)  \\ \midrule
            \multirow{4}{*}{8}   & 90.49\%            & 59.55\%            & 82.14\%            & 91.47\%            & 47.72\%             & 92.01\%             & 91.45\%            & 45.97\%             & 86.89\%             \\
                                 & ($2^{-18}\sim2^{-3}$) & ($2^{-16}\sim2^{-1}$) & ($2^{-18}\sim2^{-3}$) & ($2^{-22}\sim2^{-7}$) & ($2^{-22}\sim2^{-7}$)  & ($2^{-16}\sim2^{-1}$)  & ($2^{-20}\sim2^{-5}$) & ($2^{-20}\sim2^{-5}$)  & ($2^{-20}\sim2^{-5}$)  \\
                                 & 90.46\%            & 56.76\%            & 83.12\%            & 91.47\%            & 40.02\%             & 91.96\%             & 90.63\%            & 21.94\%             & 68.56\%             \\
                                 & ($2^{-10}\sim2^{-3}$) & ($2^{-10}\sim2^{-3}$) & ($2^{-12}\sim2^{-5}$) & ($2^{-14}\sim2^{-7}$) & ($2^{-16}\sim2^{-9}$)  & ($2^{-14}\sim2^{-7}$)  & ($2^{-14}\sim2^{-7}$) & ($2^{-16}\sim2^{-9}$)  & ($2^{-14}\sim2^{-7}$)  \\ \midrule
            Max. Segment         & \multicolumn{9}{c}{APoT-PWLF (Accuracy/Exponent Range)}                                                                                                                                        \\ \midrule
            \multirow{4}{*}{6}   & 91.04\%            & 86.89\%            & 85.16\%            & 92.12\%            & 76.92\%             & 91.96\%             & 91.79\%            & 80.14\%             & 88.45\%             \\
                                 & ($2^{-18}\sim2^{-3}$) & ($2^{-16}\sim2^{-1}$) & ($2^{-20}\sim2^{-5}$) & ($2^{-18}\sim2^{-3}$) & ($2^{-22}\sim2^{-7}$)  & ($2^{-20}\sim2^{-5}$)  & ($2^{-20}\sim2^{-5}$) & ($2^{-18}\sim2^{-3}$)  & ($2^{-22}\sim2^{-7}$)  \\
                                 & 90.93\%            & 79.28\%            & 86.04\%            & 92.09\%            & 47.46\%             & 89.57\%             & 91.71\%            & 40.81\%             & 87.64\%             \\
                                 & ($2^{-10}\sim2^{-3}$) & ($2^{-10}\sim2^{-3}$) & ($2^{-12}\sim2^{-5}$) & ($2^{-14}\sim2^{-7}$) & ($2^{-16}\sim2^{-9}$)  & ($2^{-18}\sim2^{-11}$) & ($2^{-14}\sim2^{-7}$) & ($2^{-16}\sim2^{-9}$)  & ($2^{-14}\sim2^{-7}$)  \\ \midrule
            \multirow{4}{*}{8}   & 91.14\%            & 86.16\%            & 85.78\%            & 92.09\%            & 85.92\%             & 92.42\%             & 91.94\%            & 75.49\%             & 89.26\%             \\
                                 & ($2^{-18}\sim2^{-3}$) & ($2^{-16}\sim2^{-1}$) & ($2^{-18}\sim2^{-3}$) & ($2^{-18}\sim2^{-3}$) & ($2^{-22}\sim2^{-7}$)  & ($2^{-20}\sim2^{-5}$)  & ($2^{-22}\sim2^{-7}$) & ($2^{-20}\sim2^{-5}$)  & ($2^{-22}\sim2^{-7}$)  \\
                                 & 90.94\%            & 77.83\%            & 85.82\%            & 92.04\%            & 75.46\%             & 92.38\%             & 91.83\%            & 23.97\%             & 85.14\%             \\
                                 & ($2^{-10}\sim2^{-3}$) & ($2^{-10}\sim2^{-3}$) & ($2^{-12}\sim2^{-5}$) & ($2^{-14}\sim2^{-7}$) & ($2^{-16}\sim2^{-9}$)  & ($2^{-16}\sim2^{-9}$)  & ($2^{-14}\sim2^{-7}$) & ($2^{-18}\sim2^{-11}$) & ($2^{-16}\sim2^{-9}$)  \\ \bottomrule
        \end{tabular}
    }
    \vspace{-5pt}
    \label{cifar10_results}
\end{table*}

\begin{table*}[t]
    \centering
    \caption{Design Space Exploration for APoT-PWLF Approximation based on Original QNN and PWLF Models with ImageNet~\cite{imagenet-dataset} and ResNet-18~\cite{resnet} Architecture}
    \resizebox{\linewidth}{!}
    {
        \begin{tabular}{ccccccccccccc}
            \toprule
            \multicolumn{13}{c}{Experiment Result}                                                                                                                                         \\ \midrule
            Precision            & \multicolumn{6}{c}{8-bit}                                                             & \multicolumn{6}{c}{Mixed-Precision}                                                   \\
            Activation           & \multicolumn{2}{c}{ReLU}                  & \multicolumn{2}{c}{ReLU+SiLU}             & \multicolumn{2}{c}{ReLU+GELU}             & \multicolumn{2}{c}{ReLU}                  & \multicolumn{2}{c}{ReLU+SiLU}             & \multicolumn{2}{c}{ReLU+GELU}             \\
            \multicolumn{1}{l}{} & Top-1                & Top-5              & Top-1               & Top-5               & Top-1               & Top-5               & Top-1               & Top-5               & Top-1                & Top-5              & Top-1               & Top-5               \\
            Original             & 67.07\%              & 87.23\%            & 67.44\%             & 87.23\%             & 66.79\%             & 87.02\%             & 65.41\%              & 86.29\%            & 65.26\%             & 86.10\%             & 65.18\%             & 86.01\%             \\ \midrule
            Max. Segment              & \multicolumn{12}{c}{PWLF}                                                                                                                                                      \\ \midrule
            6                    & 66.71\%              & 86.96\%            & 66.50\%             & 86.66\%             & 65.88\%             & 86.64\%             & 62.63\%              & 84.13\%            & 56.78\%             & 79.82\%             & 51.52\%             & 76.48\%             \\
            8                    & 66.69\%              & 87.01\%            & 67.17\%             & 87.06\%             & 66.63\%             & 86.90\%             & 64.23\%              & 85.47\%            & 62.20\%             & 84.29\%             & 59.33\%             & 82.00\%             \\ \midrule
            Max. Segment              & \multicolumn{12}{c}{APoT-PWLF (Accuracy/Exponent Range)}                                                                                                                       \\ \midrule
            \multirow{4}{*}{6}   & 66.67\%              & 86.97\%            & 66.47\%             & 86.64\%             & 65.92\%             & 86.43\%             & 62.48\%              & 84.11\%            & 56.84\%             & 79.75\%             & 51.96\%             & 76.57\%             \\
                                 & \multicolumn{2}{c}{($2^{-22}\sim2^{-7}$)} & \multicolumn{2}{c}{($2^{-22}\sim2^{-7}$)} & \multicolumn{2}{c}{($2^{-22}\sim2^{-7}$)} & \multicolumn{2}{c}{($2^{-20}\sim2^{-5}$)} & \multicolumn{2}{c}{($2^{-18}\sim2^{-3}$)} & \multicolumn{2}{c}{($2^{-20}\sim2^{-5}$)} \\
                                 & 66.65\%              & 86.92\%            & 64.35\%             & 85.46\%             & 64.57\%             & 85.63\%             & 59.55\%              & 82.09\%            & 46.65\%             & 71.43\%             & 49.76\%             & 75.02\%             \\
                                 & \multicolumn{2}{c}{($2^{-16}\sim2^{-9}$)} & \multicolumn{2}{c}{($2^{-16}\sim2^{-9}$)} & \multicolumn{2}{c}{($2^{-16}\sim2^{-9}$)} & \multicolumn{2}{c}{($2^{-14}\sim2^{-7}$)} & \multicolumn{2}{c}{($2^{-12}\sim2^{-5}$)} & \multicolumn{2}{c}{($2^{-14}\sim2^{-7}$)} \\ \midrule
            \multirow{4}{*}{8}   & 66.75\%              & 87.00\%            & 67.22\%             & 87.04\%             & 66.70\%             & 86.99\%             & 64.28\%              & 85.48\%            & 62.56\%             & 84.39\%             & 59.65\%             & 82.19\%             \\
                                 & \multicolumn{2}{c}{($2^{-22}\sim2^{-7}$)} & \multicolumn{2}{c}{($2^{-18}\sim2^{-3}$)} & \multicolumn{2}{c}{($2^{-16}\sim2^{-1}$)} & \multicolumn{2}{c}{($2^{-18}\sim2^{-3}$)} & \multicolumn{2}{c}{($2^{-22}\sim2^{-7}$)} & \multicolumn{2}{c}{($2^{-20}\sim2^{-5}$)} \\
                                 & 66.71\%              & 86.92\%            & 65.86\%             & 86.35\%             & 66.28\%             & 86.87\%             & 63.41\%              & 84.95\%            & 61.45\%             & 83.54\%             & 58.51\%             & 81.31\%             \\
                                 & \multicolumn{2}{c}{($2^{-16}\sim2^{-9}$)} & \multicolumn{2}{c}{($2^{-14}\sim2^{-7}$)} & \multicolumn{2}{c}{($2^{-16}\sim2^{-9}$)} & \multicolumn{2}{c}{($2^{-14}\sim2^{-7}$)} & \multicolumn{2}{c}{($2^{-14}\sim2^{-7}$)} & \multicolumn{2}{c}{($2^{-14}\sim2^{-7}$)} \\ \bottomrule       
        \end{tabular}
    }    
    \vspace{-5pt}
    \label{imagenet_results}
\end{table*}

In the early stage of this work, we adopted the \emph{pwlf}~\cite{pwlf} library to fit PWLF representations before approximating their slopes with PoT or APoT values. However, \emph{pwlf} searches for breakpoint locations using differential evolution, making channel-wise fitting prohibitively slow for larger QNNs. On our AMD EPYC 7513 32-core server, fitting one nonlinear function with 1,000 sampled points takes approximately four minutes. As an example, ResNet-26 requires approximately 4,904 channel-wise activation functions to be fitted. As a result, processing a single model configuration would take an estimated 13.6 days, making repeated design-space exploration impractical. 

Therefore, to adopt our PoT and APoT PWLF approximation algorithm to larger networks and datasets, like ResNet-18 and ImageNet, we developed a fast greedy piecewise linear fitting algorithm as shown in~\autoref{alg:greedy_pwlf}. This algorithm employs a greedy integer-aware breakpoint selection strategy for PWLF. It starts from a single segment spanning the entire input range and iteratively refines the approximation. For each current segment, the point with the maximum vertical distance between the target nonlinear function and the chord connecting the segment endpoints is identified. This point is then rounded to the nearest integer to satisfy the integer breakpoint constraint required by hardware implementation. A candidate breakpoint is accepted only if it lies inside the segment and provides an improvement larger than the minimum threshold $\epsilon$. Among all valid candidates, the one with the largest distance is greedily selected to split the corresponding segment into two subsegments. The process continues until the target number of segments is reached or no valid breakpoint can provide sufficient improvement. This fast greedy PWLF algorithm can fit all nonlinear activation functions in ResNet-18 in less than 1 mins with 8 segments. 

\begin{algorithm}[t]
\caption{Greedy Integer-Aware PWLF Breakpoint Selection}
\label{alg:greedy_pwlf}
\KwIn{sampled points $(x_i,y_i)$, target segments $K$, min improvement $\epsilon$}
\KwOut{breakpoint set $\mathcal{B}$}

sort $(x_i,y_i)$ by $x$\;
$\mathcal{B}\gets\emptyset$, $\mathcal{S}\gets\{(x_1,x_N)\}$\;

\While{$|\mathcal{B}| < K-1$}{
    $\mathcal{P}\gets\emptyset$\;
    \ForEach{$(a,b)\in\mathcal{S}$}{
        find $x^\star=\arg\max$ distance to chord $(a,b)$\;
        $\hat{x}\gets\mathrm{round}(x^\star)$\;
        \If{$a<\hat{x}<b$ and $\mathrm{dist}(x^\star)>\epsilon$}{
            add $(\mathrm{dist}(x^\star),\hat{x},(a,b))$ to $\mathcal{P}$\;
        }
    }
    \If{$\mathcal{P}=\emptyset$}{
        \textbf{break}\;
    }
    select valid $(\hat{x},(a,b))$ from $\mathcal{P}$ with largest distance\;
    $\mathcal{B}\gets\mathcal{B}\cup\{\hat{x}\}$\;
    replace $(a,b)$ with $(a,\hat{x})$ and $(\hat{x},b)$ in $\mathcal{S}$\;
}
\Return{$\mathcal{B}$}\;
\end{algorithm}

Based on this algorithm, we conduct a design-space exploration (DSE) for PoT/APoT-PWLF approximation on VGG16/CIFAR-10 and ResNet-18/ImageNet to find a suitable hardware design reference. For VGG16, we evaluated our PWLF approximation under 4/8-bit unified-precision quantization and 2/4/8-bit mixed-precision quantization. For ResNet-18, we evaluated its 8-bit unified-precision quantization and 2/4/8-bit mixed-precision quantization models. For the mixed-precision quantization model, we adopted the same precision across all layers in one stage of VGG16 and ResNet-18, with 8/4/2/4/8-bit precision for each stage and the fully connected layer. For VGG16 models, we replace all activations with ReLU/Sigmoid/SiLU. And for ResNet-18 models, we evaluated the models with ReLU activation, ReLU/SiLU-mixed, and ReLU/GELU-mixed models (The layers in the fourth stage of the ResNet-18 model adopted SiLU/GELU activation functions). \autoref{cifar10_results} and \autoref{imagenet_results} listed our approximate results on CIFAR-10 and ImageNet dataset. We thoroughly explored our approximate schemes based on our PWLF function with 6/8 segments and continuous 8/16 exponents of $2^n$. Because the PoT-PWLF has higher accuracy loss than APoT-PWLF for ResNet-18 and ImageNet, we only list the APoT-PWLF approximation results in \autoref{imagenet_results}. 

The experimental results show that no single configuration is optimal for all models. Compared with ReLU, Sigmoid, SiLU, and GELU are more sensitive to PWLF approximation. In most cases, increasing the segment number or the number of allowable $2^n$ values improves accuracy. More segments enable PWLF to fit the original nonlinear function more accurately, while more $2^n$ values provide more choices for PoT and APoT slope approximation.

The results also show that the accuracy of the PWLF model mainly determines the upper limit of PoT-PWLF and APoT-PWLF. When PWLF maintains the original model accuracy, APoT-PWLF can also recover almost all the accuracy of the original QNN. For example, the accuracy difference between APoT-PWLF and the original QNN is no more than 0.17 percentage points for all VGG16 ReLU models. For the three 8-bit ResNet-18 models, APoT-PWLF limits the Top-1 and Top-5 accuracy losses to $0.09\sim0.32$ and $0.03\sim0.23$ percentage points, respectively. However, if PWLF itself introduces a large accuracy loss, APoT approximation cannot recover this loss. For example, with eight segments, the mixed-precision ReLU/GELU model based on PWLF and APoT-PWLF only achieves $59.33\%/82.00\%$ and $59.65\%/82.19\%$ Top-1/Top-5 accuracy, respectively, compared with $65.18\%/86.01\%$ for the original QNN. Therefore, we further increased the maximum segment number to 20. The PWLF model then reaches $64.65\%/85.71\%$, while APoT-PWLF with the exponent range $2^{-18}\sim2^{-3}$ reaches $64.83\%/85.74\%$, reducing the remaining loss relative to the original QNN to only 0.35/0.27 percentage points. This result suggests that difficult nonlinear functions can recover the original model accuracy by using more segments, and such configurations can be supported by the serialized GRAU design when additional processing cycles are acceptable.

\begin{figure}[t]
    \centering
    \begin{minipage}{\columnwidth}
        \centerline{\includegraphics[width=1\columnwidth]{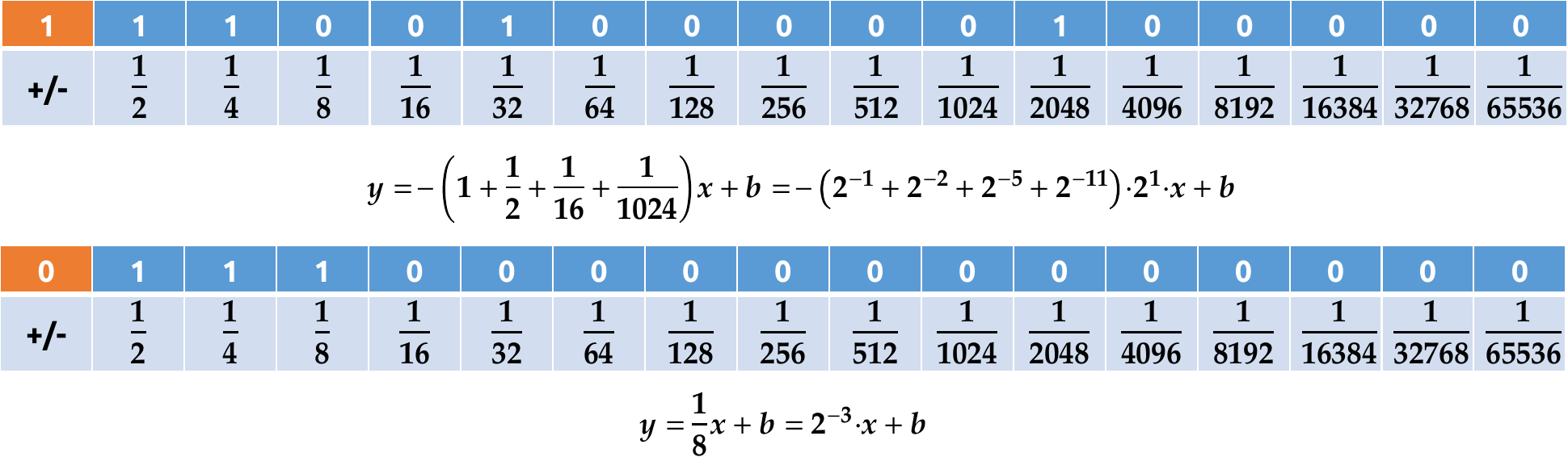}}
    \end{minipage}
    \caption{Encoding of segment slopes for PoT-PWLF (down) and APoT-PWLF (up) approximation} \vspace{5pt}
    \label{segment_setting}
    \begin{minipage}{\columnwidth}
        \centerline{\includegraphics[width=1\columnwidth]{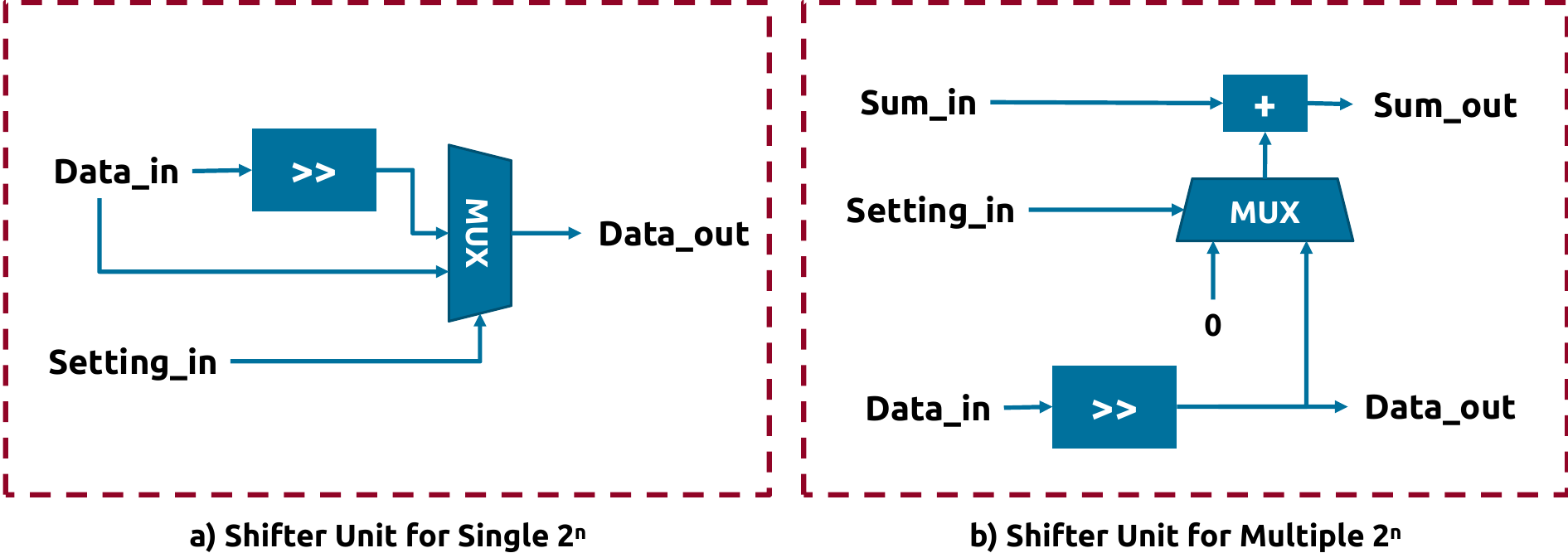}}
    \end{minipage}
    \caption{Shifter unit design in hardware for PoT-PWLF (a) and APoT-PWLF (b) approximation}
    \label{pwlf_hardware}
\end{figure}

\subsection{Hardware Implementation of GRAU}

As shown in~\autoref{pwlf_hardware}, we implemented different 1-bit right shifter unit pipelines for PoT-PWLF and APoT-PWLF functions: 

For PoT-PWLF, the shifter unit in~\autoref{pwlf_hardware} (a) loads the input data and, according to the 1-bit setting input, decides whether to pass the 1-bit right-shifted data or the original input data to the next shifter unit. For APoT-PWLF, the shifter unit in~\autoref{pwlf_hardware} (b) loads the input data and a sum output from the prior shifter unit. After applying the 1-bit right shift to the input data, according to the 1-bit setting input, this shifter unit will determine whether to add the right-shifted data to the loaded sum output. Then, it will transfer the 1-bit right-shifted input and the sum result to the next shifter unit. 

As a result, if we can implement an 8/16 right shifter unit and connect them as a pipeline, we can compute the results of the inputs times PoT or APoT slopes, consisting of 8/16 available $2^n$. To control these shifter units, as shown in~\autoref{segment_setting}, we defined the shifter control encoding format, which is the~\emph{Setting In} signals in each shifter unit of~\autoref{pwlf_hardware}.~\autoref{segment_setting} shows the 17-bit shifter control encoding for 16 shifter units in PoT and APoT approximation with the power range of $[-16, -1]$. Its first bit is the sign bit. 

\begin{figure}[t]
    \centering
    \begin{minipage}{\columnwidth}
        \centerline{\includegraphics[width=1\columnwidth]{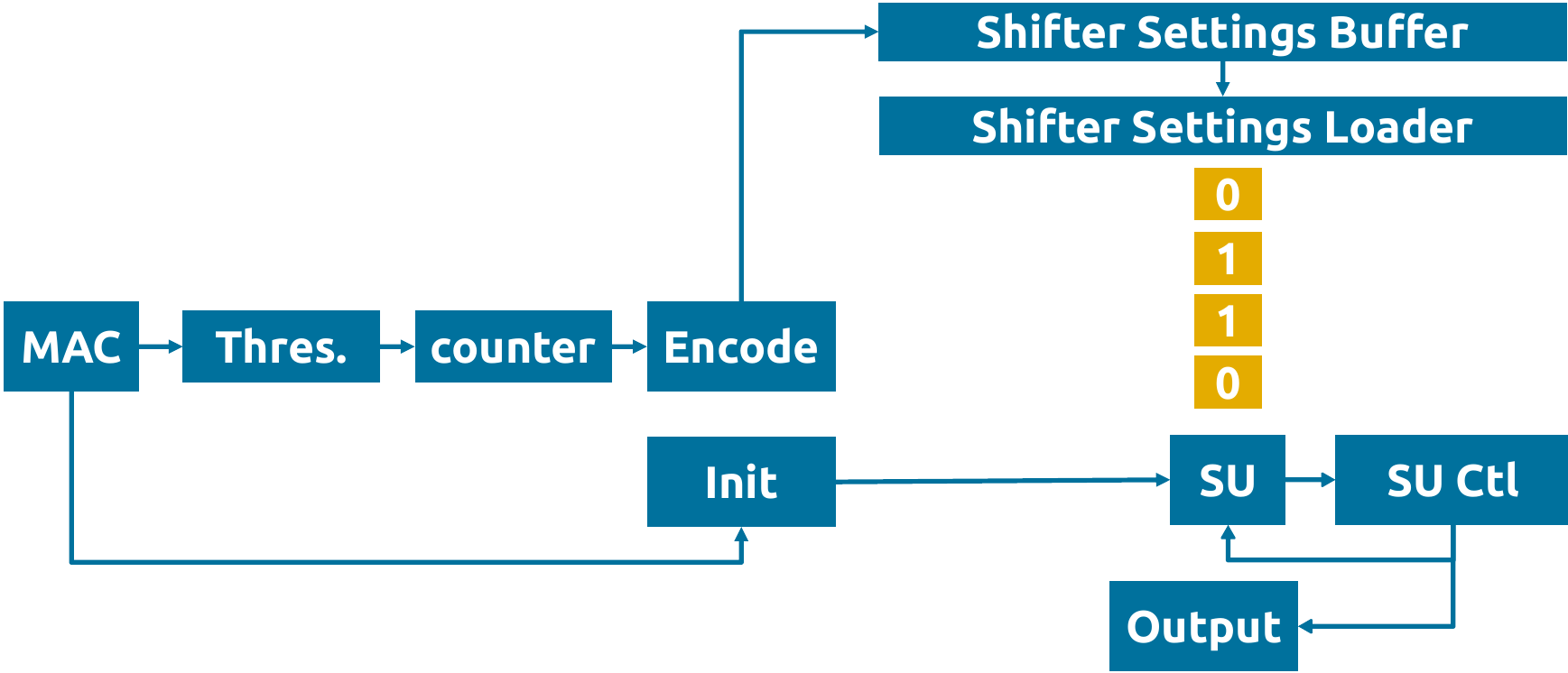}}
    \end{minipage}
    \caption{Serialized hardware implementation of Generic Activation Unit PoT-PWLF and APoT-PWLF shifter unit} \vspace{5pt}
    \label{serial_hardware}
    \begin{minipage}{\columnwidth}
        \centerline{\includegraphics[width=1\columnwidth]{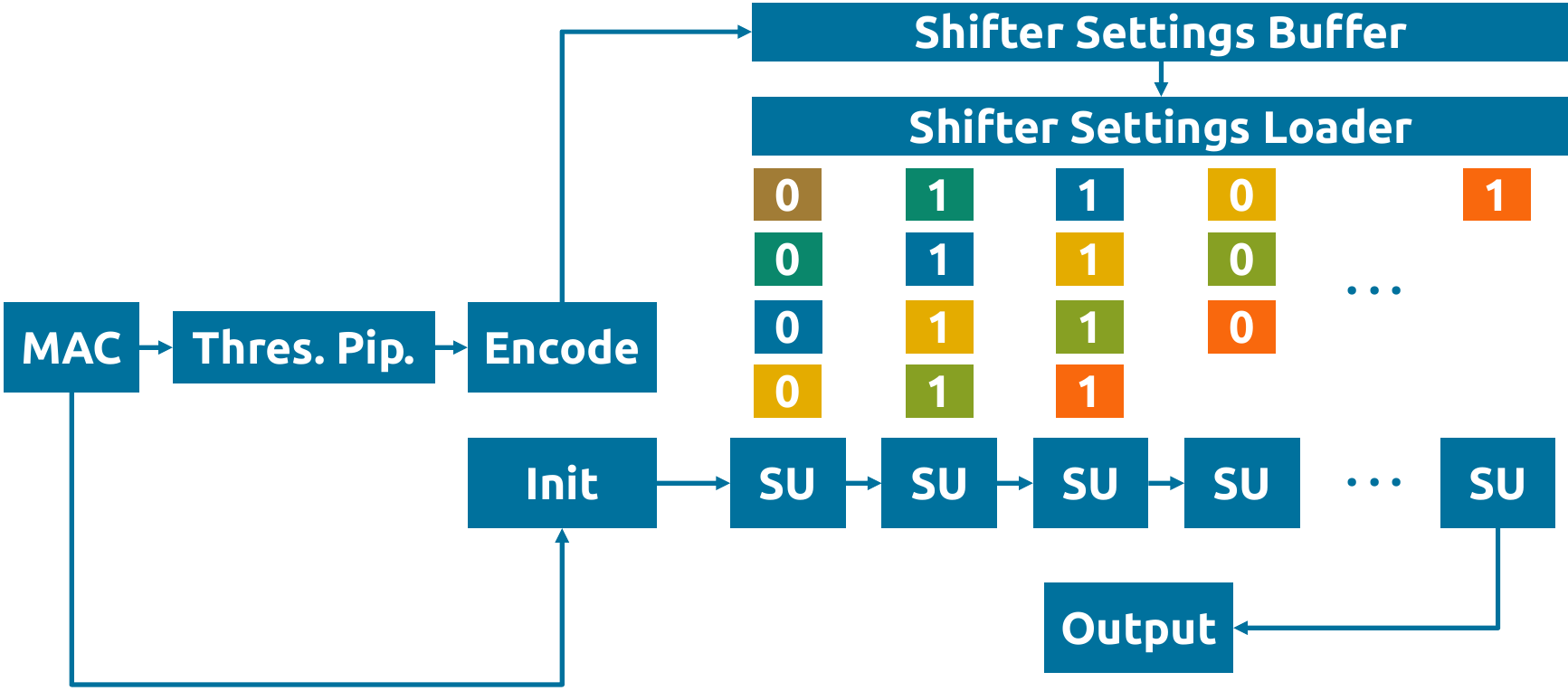}}
    \end{minipage}
    \caption{Pipelined hardware implementation of Generic Activation Unit PoT-PWLF and APoT-PWLF shifter unit}
    \label{pip_hardware}
\end{figure}

For the PoT-PWLF shifter computation in~\autoref{segment_setting} (down), because it only supports the format of a single $2^n$, the setting should consist of multiple consecutive "1" and consecutive "0". For instance, if the slope is $\frac{1}{8}$, the digits from  $\frac{1}{2}$ to $\frac{1}{8}$ should be "1". The result from the last shifter unit is equal to the original input divided by 8 (right-shifting 3 bits). For APoT-PWLF shifter computation in~\autoref{segment_setting} (up), if the slope is $1+\frac{1}{2}+\frac{1}{16}+\frac{1}{1024}$, the \emph{Setting In} digits for $\frac{1}{2}$, $\frac{1}{4}$, $\frac{1}{32}$, and $\frac{1}{2048}$ should be "1".  Then, the 1-bit pre-left-shifted input (Equal to $\times2$) will be executed 1-bit right-shifted in every shifter unit, and the shifted data in 1, $\frac{1}{2}$, $\frac{1}{16}$, and $\frac{1}{1024}$ units will be added into the sum. 

If all shifter encoding bits are 0, it means the slope is 0.

Based on evaluation results in~\autoref{cifar10_results} and \autoref{imagenet_results}, considering we limit the exponent range of slopes in PoT-PWLF and APoT-PWLF functions to 8/16 continuous $2^n$, and $n$ is always negative, we give a pre-shifting, which means dividing by $2^{-m}$ to every input. Then, the shifter pipeline will only need to process the continuous right shifting in $[2^{-8}, 2^{-1}]$ and $[2^{-16}, 2^{-1}]$. Therefore, in the real hardware implementation, all shifting encodings are negative. 

Therefore, based on the above-mentioned principle, we implemented two different GRAU architectures for PoT-PWLF and APoT-PWLF shifter units. \autoref{serial_hardware} shows the serialized architecture of GRAU. It only implements one shifter unit and reuses it for different slope approximations.~\autoref{pip_hardware} shows the pipelined architecture of GRAU, which consists of a shifter unit pipeline. 

\begin{table*}[t]
    \centering
    \caption{Hardware Results of Multi-Threshold, PoT-PWLF, and APoT-PWLF Activation Units}
    \vspace{-5pt}
    \resizebox{\linewidth}{!}
    {
        \begin{tabular}{ccccccccccccccc}
            \toprule
            \multirow{2}{*}{Activation Unit} & \multirow{2}{*}{Design}    & Segment & Exponent & \multirow{2}{*}{LUT} & \multirow{2}{*}{FF} & \multirow{2}{*}{Frequency} & Total & Dynamic  & Power-Delay & Area-Delay & \multicolumn{4}{c}{Pipeline Depth} \\
                                             &                            & Number  & Number    &                      &                     &                            & Delay (ns)        & Power (W)           & Products    & Products   & 1-bit   & 2-bit  & 4-bit  & 8-bit  \\ \midrule
            \multirow{2}{*}{Multi-Threshold} & Pipelined                  & -       & -        & 10206                & 18568               & 200MHz                     & 2.848       & 0.129         & 0.3674      & 29066.688  & 1       & 3      & 15     & 255    \\
                                             & Serialization              & -       & -        & 2796                 & 8264                & 100MHz                     & 5.777       & 0.032         & 0.1849      & 16152.492  & -       & -      & -      & -      \\ \midrule
            \multirow{5}{*}{PoT-PWLF}        & \multirow{4}{*}{Pipelined} & 6       & 8        & 408                  & 675                 & \multirow{5}{*}{250MHz}    & 1.658       & 0.013         & 0.0216      & 676.464    & 1       & 3      & 16     & 16     \\
                                             &                            & 6       & 16       & 647                  & 1007                &                            & 1.570       & 0.015         & 0.0236      & 1015.790   & 1       & 3      & 24     & 24     \\
                                             &                            & 8       & 8        & 507                  & 854                 &                            & 1.811       & 0.013         & 0.0235      & 918.177    & 1       & 3      & 18     & 18     \\
                                             &                            & 8       & 16       & 755                  & 1202                &                            & 1.655       & 0.015         & 0.0248      & 1249.525   & 1       & 3      & 26     & 26     \\
                                             & Serialization              & -       & -        & 270                  & 456                 &                            & 2.338       & 0.012         & 0.0281      & 631.260    & -       & -      & -      & -      \\ \midrule
            \multirow{5}{*}{APoT-PWLF}       & \multirow{4}{*}{Pipelined} & 6       & 8        & 458                  & 709                 & \multirow{5}{*}{250MHz}    & 1.858       & 0.013         & 0.0242      & 850.964    & 1       & 3      & 16     & 16     \\
                                             &                            & 6       & 16       & 786                  & 1097                &                            & 1.946       & 0.016         & 0.0311      & 1529.556   & 1       & 3      & 24     & 24     \\
                                             &                            & 8       & 8        & 558                  & 888                 &                            & 1.608       & 0.013         & 0.0209      & 897.264    & 1       & 3      & 18     & 18     \\
                                             &                            & 8       & 16       & 895                  & 1292                &                            & 1.775       & 0.017         & 0.0302      & 1588.625   & 1       & 3      & 26     & 26     \\
                                             & Serialization              & -       & -        & 283                  & 463                 &                            & 2.352       & 0.011         & 0.0259      & 665.616    & -       & -      & -      & -     \\ \bottomrule
        \end{tabular}
    }
    \vspace{-5pt}
    \label{hardware_results}
\end{table*}

The integer outputs of the MAC will be loaded into the thresholds first to determine which segment they belong to. As shown in~\autoref{cifar10_results} or \autoref{imagenet_results}, for a PWLF function with 6 segments, if we consider the inputs, which are out of the range of the function approximation, belong to the first and last segments, we only need to implement 5 thresholds to classify the inputs. After obtaining the index of the segment by thresholds, GRAU loads the shifter unit setting from the setting buffer, which is a look-up table, and passes it to the setting loader to regenerate the setting encoded in the correct format for both pipelined and serialized GRAU designs. The MAC output will be transmitted to the initial module simultaneously to apply the pre-right shift. After both input and shifter settings are ready for processing, they will be loaded into the serialized shifter unit or shifter unit pipeline to compute the products between the input and the slope, approximated in the PoT or APoT form. The final result will apply the sign bit and bias to complete the piecewise linear approximation for the activation unit. Therefore, based on the above-mentioned designs, if we reload the value of thresholds and shifter settings, GRAU can be reconfigured to approximate different nonlinear functions for the QNN acceleration on hardware. 

\section{Evaluation}
\label{evaluation}

To evaluate this work, we implemented 12 different activation unit instances on hardware based on \emph{Multi-Threshold} (MT) architecture and PoT/APoT-PWLF-based GRAU architectures. Three of these 12 instances are implemented as a serialized structure, and the others are pipelined structures. Considering that the \emph{Multi-Threshold} activation unit can reconfigure thresholds for different activation functions, which can also be considered a kind of generic reconfigurable activation unit, we include it in this evaluation as a baseline. This baseline follows the official FINN-R implementation\cite{Blott2018}, where each output precision level requires $2^n-1$ thresholds after BN folding. This behavior is inherent to the MT paradigm rather than a design choice, and is well documented in prior FINN literature. We optimized it as a pipelined structure consisting of 255 threshold units connected in a pipeline, and a serialized structure that implements only one reusable threshold with 255 threshold registers. 

Therefore, as shown in~\autoref{hardware_results}, we synthesized and implemented these instances using \emph{Vivado} on the \emph{Ultra96-V2} FPGA platform. Based on the post-implementation timing simulation, we evaluated their power, critical path delay, \emph{Area-Delay-Product} (ADP), and \emph{Power-Delay-Product} (PDP).

\subsubsection{Hardware Resource Consumption} \autoref{hardware_results} shows that both PoT-PWLF and APoT-PWLF activation units are much more LUT-efficient than the MT unit. Compared with the pipelined MT design requiring 10206 LUTs, the proposed pipelined PoT/APoT-PWLF units only use $408\sim755$ and $458\sim895$ LUTs, respectively, i.e., only $4.0\%\sim7.4\%$ and $4.5\%\sim8.8\%$ of the MT cost. A similar trend also holds for the serialized designs (270/283 vs. 2796 LUTs). This advantage mainly comes from the fact that the PWLF-based design scales with the number of segments and exponent candidates, whereas the MT unit scales with maximum output precision, as also reflected by its much deeper 255-threshold pipeline. Moreover, within the proposed design, increasing the segment number is generally more LUT-efficient than increasing the exponent number. For example, in PoT-PWLF, increasing the exponent number from 8 to 16 raises the LUT count from 408 to 647 for the 6-segment case, while increasing the segment number from 6 to 8 with 8 exponents only raises the LUT count from 408 to 507. Similar trends can also be observed for APoT-PWLF. Combined with the accuracy results, this suggests that allocating hardware budget to more segments is often more cost-effective than allocating it to more $2^n$ candidates, while larger exponent sets are mainly beneficial for the most difficult cases.

\subsubsection{Pipeline Depth} As shown in the pipeline depth column in~\autoref{hardware_results}, pipelined \emph{Multi-Threshold} activation unit takes 1/3/15/255 cycles to processing one input. Because our pipeline GRAU instances have one pre-right-shifting unit, 5/7 thresholds, 8/16 right shifters, one sign bit processing unit, and one bias adder, it will take $16\sim24$ cycles to complete one approximate nonlinear processing, which is a bit slower than the 4-bit quantization of the \emph{Multi-Threshold} activation unit, but it is substantially faster than the 255-cycle 8-bit MT unit. Furthermore, considering that GRAU also has thresholds, it can support the 1/2-bit \emph{Multi-Threshold} quantization with 1/3 thresholds. We implemented a bypass for our GRAU instances for 1/2-bit. Therefore, as shown in the latency column in~\autoref{hardware_results}, they take the same cycles as MT units in 1/2-bit columns.

\subsubsection{Total Delay, Power, ADP, and PDP} The post-implementation timing simulation on \emph{Vivado} reported the power, critical path delay, \emph{Area-Delay-Product} (ADP), and \emph{Power-Delay-Product} (PDP) of our 12 instances. All pipelined GRAU implementations meet the target frequency of 250 MHz, compared with 200 MHz for the pipelined MT baseline. Therefore, we can infer that our GRAU implementations can support higher clock frequencies due to their low total delay. Moreover, the lower ADP and PDP of our GRAU also show that our work has better power and design efficiency than the \emph{Multi-Threshold} activation units. 

\section{Conclusion and Further Works}
\label{concl}
 To support the multi-function, mixed-precision QNN and optimize the hardware resource consumption, we explored the \emph{\textbf{G}eneric \textbf{R}econfigurable \textbf{A}ctivation \textbf{U}nit} (GRAU) based on our fast greedy \emph{Piecewise Linear Fitting} approximated algorithm to generate \emph{Power-of-Two approximated PWLF}  function and \emph{Additive Power-of-Two approximated PWLF} function. The DSE results for the \textit{VGG16} and \textit{ResNet-18} models, using the \textit{CIFAR-10} and ImageNet datasets, show that APoT-PWLF can closely recover the corresponding PWLF accuracy, while its accuracy relative to the original QNN is mainly limited by the PWLF segment number. In the most difficult mixed-precision ReLU/GELU case, increasing the segment number to 20 reduces the remaining Top-1/Top-5 accuracy loss to only 0.35/0.27 percentage points. Therefore, we implemented the serialized and pipelined PoT-PWLF and APoT-PWLF approximated \emph{Generic Reconfigurable Activation Unit} (GRAU) in this work. The implementation results show that our work can reduce by approximately $90\%$ or more of the LUT consumption of \emph{Multi-Threshold}-based generic activation functions, supporting higher clock frequencies with better power and hardware efficiency.

In our current design, we set a unified approximation range of $2^n$ for all layers in network models. However, our analysis of the MAC output range in each layer suggests that the input range of our GRAU unit can be between $[-10^{6}, 10^{6}]$ or $[-10^{4}, 10^{4}]$. Therefore, we will deeply explore the fine-grained approximation for the QNN models and the design space exploration of hardware based on our PWLF algorithm in the future work. 

\section*{Acknowledgments}
This work is supported by the \emph{Center for Scalable Data Analytics and Artificial Intelligence} (ScaDS.AI) \emph{Dresden/Leipzig} and \emph{Deutsche Forschungsgemeinschaft} (DFG) under the \emph{X-ReAp} project (Project number 380524764).

%
%
\renewcommand{\bibfont}{\footnotesize}
\printbibliography

@article{hendrycks2016gelu,
  title={Gaussian Error Linear Units (GELUs)},
  author={Hendrycks, Dan and Gimpel, Kevin},
  journal={arXiv preprint arXiv:1606.08415},
  year={2016}
}

@ARTICLE{ml-plac,
  author={Lyu, Fei and Xia, Yan and Mao, Zhelong and Wang, Yanxu and Wang, Yu and Luo, Yuanyong},
  journal={IEEE Transactions on Circuits and Systems I: Regular Papers}, 
  title={ML-PLAC: Multiplierless Piecewise Linear Approximation for Nonlinear Function Evaluation}, 
  year={2022},
  volume={69},
  number={4},
  pages={1546-1559},
  doi={10.1109/TCSI.2021.3133931}}

@inproceedings{resnet,
  author    = {He, Kaiming and Zhang, Xiangyu and Ren, Shaoqing and Sun, Jian},
  title     = {Deep Residual Learning for Image Recognition},
  booktitle = {Proceedings of the IEEE Conference on Computer Vision and Pattern Recognition},
  pages     = {770--778},
  year      = {2016},
  doi       = {10.1109/CVPR.2016.90}
}

@InProceedings{Umuroglu2017,
  author    = {Umuroglu, Yaman and Fraser, Nicholas J and Gambardella, Giulio and Blott, Michaela and Leong, Philip and Jahre, Magnus and Vissers, Kees},
  booktitle = {Proceedings of the 2017 ACM/SIGDA International Symposium on Field-Programmable Gate Arrays},
  title     = {FINN: A framework for fast, scalable binarized neural network inference},
  year      = {2017},
  pages     = {65--74},
}

@Article{Blott2018,
  author    = {Blott, Michaela and Preu{\ss}er, Thomas B and Fraser, Nicholas J and Gambardella, Giulio and O’brien, Kenneth and Umuroglu, Yaman and Leeser, Miriam and Vissers, Kees},
  journal   = {ACM Transactions on Reconfigurable Technology and Systems (TRETS)},
  title     = {FINN-R: An end-to-end deep-learning framework for fast exploration of quantized neural networks},
  year      = {2018},
  number    = {3},
  pages     = {1--23},
  volume    = {11},
  publisher = {ACM New York, NY, USA},
}

@inproceedings{Simonyan2014,
  author    = {Simonyan, Karen and Zisserman, Andrew},
  title     = {Very Deep Convolutional Networks for Large-Scale Image Recognition},
  booktitle = {International Conference on Learning Representations},
  year      = {2015}
}

@article{deng2012mnist,
  title={The {MNIST} database of handwritten digit images for machine learning research},
  author={Deng, Li},
  journal={IEEE Signal Processing Magazine},
  volume={29},
  number={6},
  pages={141--142},
  year={2012},
  publisher={IEEE},
  doi={10.1109/MSP.2012.2211477},
}

@mastersthesis{Krizhevsky2009,
  author = {Krizhevsky, Alex},
  title  = {Learning Multiple Layers of Features from Tiny Images},
  school = {University of Toronto},
  year   = {2009},
  url    = {https://www.cs.toronto.edu/~kriz/learning-features-2009-TR.pdf}
}

@software{brevitas,
  author       = {Alessandro Pappalardo},
  title        = {Xilinx/brevitas},
  year         = {2023},
  publisher    = {Zenodo},
  doi          = {10.5281/zenodo.3333552},
  url          = {https://doi.org/10.5281/zenodo.3333552}
}

@INPROCEEDINGS{imagenet-dataset,
  author={Deng, Jia and Dong, Wei and Socher, Richard and Li, Li-Jia and Kai Li and Li Fei-Fei},
  booktitle={2009 IEEE Conference on Computer Vision and Pattern Recognition}, 
  title={ImageNet: A large-scale hierarchical image database}, 
  year={2009},
  volume={},
  number={},
  pages={248-255},
  doi={10.1109/CVPR.2009.5206848}}

@INPROCEEDINGS{bitsys2,
  author={Liu, Yuhao and Ullah, Salim and Kumar, Akash},
  booktitle={2025 26th International Symposium on Quality Electronic Design (ISQED)}, 
  title={Bitwise Systolic Array Architecture for Runtime-Reconfigurable Multi-Precision Quantized Multiplication on Hardware Accelerators}, 
  year={2025},
  volume={},
  number={},
  pages={1-9},
  doi={10.1109/ISQED65160.2025.11014376}}

@article{SiLU,
  title={Sigmoid-weighted linear units for neural network function approximation in reinforcement learning},
  author={Elfwing, Stefan and Uchibe, Eiji and Doya, Kenji},
  journal={Neural networks},
  volume={107},
  pages={3--11},
  year={2018},
  publisher={Elsevier},
  doi={10.1016/j.neunet.2017.12.012},
}

@inproceedings{Nguyen2020,
author = {Nguyen, Vantruong and Cai, Jueping and Wei, Linyu},
title = {Low Complexity Sigmoid Function Implementation Using Probability-Based Piecewise Linear Function},
year = {2020},
isbn = {9781450372619},
publisher = {Association for Computing Machinery},
address = {New York, NY, USA},
url = {https://doi.org/10.1145/3377713.3377769},
doi = {10.1145/3377713.3377769},
booktitle = {Proceedings of the 2019 2nd International Conference on Algorithms, Computing and Artificial Intelligence},
pages = {236-–241},
numpages = {6},
series = {ACAI '19}
}

@INPROCEEDINGS{Tsmots2019,
  author={Tsmots, Ivan and Skorokhoda, Oleksa and Rabyk, Vasyl},
  booktitle={2019 IEEE 15th International Conference on the Experience of Designing and Application of CAD Systems (CADSM)}, 
  title={Hardware Implementation of Sigmoid Activation Functions using FPGA}, 
  year={2019},
  volume={},
  number={},
  pages={34-38},
  doi={10.1109/CADSM.2019.8779253}}

@Article{Li2022,
AUTHOR = {Li, Zerun and Zhang, Yang and Sui, Bingcai and Xing, Zuocheng and Wang, Qinglin},
TITLE = {FPGA Implementation for the Sigmoid with Piecewise Linear Fitting Method Based on Curvature Analysis},
JOURNAL = {Electronics},
VOLUME = {11},
YEAR = {2022},
NUMBER = {9},
pages = {1365},
URL = {https://www.mdpi.com/2079-9292/11/9/1365},
ISSN = {2079-9292},
DOI = {10.3390/electronics11091365}
}

@article{Liu2023_2,
title = {Cost effective Tanh activation function circuits based on fast piecewise linear logic},
journal = {Microelectronics Journal},
volume = {138},
pages = {105821},
year = {2023},
issn = {1879-2391},
doi = {10.1016/j.mejo.2023.105821},
url = {https://www.sciencedirect.com/science/article/pii/S0026269223001349},
author = {Kezhu Liu and Weiwei Shi and Canji Huang and Dongdong Zeng}
}

@INPROCEEDINGS{Bouguezzi,
  author={Bouguezzi, Safa and Faiedh, Hassene and Souani, Chokri},
  booktitle={2021 18th International Multi-Conference on Systems, Signals, and Devices (SSD)}, 
  title={Hardware Implementation of Tanh Exponential Activation Function using FPGA}, 
  year={2021},
  volume={},
  number={},
  pages={1020-1025},
  doi={10.1109/SSD52085.2021.9429506}}

@INPROCEEDINGS{Piazza1993,
  author={Piazza, F. and Uncini, A. and Zenobi, M.},
  booktitle={Proceedings of 1993 International Conference on Neural Networks (IJCNN-93-Nagoya, Japan)}, 
  title={Neural networks with digital LUT activation functions}, 
  year={1993},
  volume={2},
  pages={1401--1404},
  doi={10.1109/IJCNN.1993.716806}}

@INPROCEEDINGS{Kumar2010,
  author = {Meher, Pramod Kumar},
  booktitle={2010 18th IEEE/IFIP International Conference on VLSI and System-on-Chip}, 
  title={An optimized lookup-table for the evaluation of sigmoid function for artificial neural networks}, 
  year={2010},
  volume={},
  number={},
  pages={91-95},
  doi={10.1109/VLSISOC.2010.5642617}}

@INPROCEEDINGS{Pogiri2022,
  author={Pogiri, Revathi and Ari, Samit and Mahapatra, K K},
  booktitle={2022 IEEE International Symposium on Smart Electronic Systems (iSES)}, 
  title={Design and FPGA Implementation of the LUT based Sigmoid Function for DNN Applications}, 
  year={2022},
  volume={},
  number={},
  pages={410-413},
  doi={10.1109/iSES54909.2022.00090}}

@Manual{pwlf,
author = {Jekel, Charles F. and Venter, Gerhard},
title = {{pwlf:} A Python Library for Fitting 1D Continuous Piecewise Linear Functions},
year = {2019},
url = {https://github.com/cjekel/piecewise_linear_fit_py}
}
\end{document}